\documentclass[aps,pra,reprint,floatfix]{revtex4-2}

\usepackage{custom_aps}
\PassOptionsToPackage{full}{textcomp}	% required for newtx
\usepackage{newtxtext,newtxmath}		% fonts
\usepackage{bm}							% artificial bold math (for symbols that don't have real bold)
\crefname{section}{Sec.}{Secs.}			% correct references for sections
\crefname{figure}{Fig.}{Figs.}			% correct references for figures
\Crefname{figure}{Figure}{Figures}		% correct references for figures
\crefrangelabelformat{subfigure}{#3#1#4--#5(\crefstripprefix{#1}{#2}#6}	% required reference format

\usepackage[caption=false]{subfig}		% subcaption incompatible with RevTeX
\newcommand{\phantomsubfloat}[1]{
	\captionsetup[subfigure]{labelformat=empty,aboveskip=0pt,belowskip=0pt}	% apply caption setup only temporarily
	\subfloat[][]{#1}
}

\pdfoutput=1

\definecolor{linkcolour}{rgb}{0.02,0.12,0.3}
\hypersetup{colorlinks=true,citecolor=linkcolour,linkcolor=linkcolour,urlcolor=linkcolour}

\newcommand{\ra}[1]{\renewcommand{\arraystretch}{#1}}

\DeclareSIUnit{\gauss}{G}

\begin{document}
\title{\texorpdfstring{Characterisation of three-body loss in \ce{^166Er}\\and optimised production of large Bose--Einstein condensates}{Characterisation of three-body loss in 166Er and optimised production of large Bose--Einstein condensates}}

\author{Milan Krstaji\'{c}}
\thanks{M.~K., P.~J.\ and J.~K.\ contributed equally to this work.}
\author{Péter Juhász}
\thanks{M.~K., P.~J.\ and J.~K.\ contributed equally to this work.}
\author{Ji\v{r}\'{i} Ku\v{c}era}
\thanks{M.~K., P.~J.\ and J.~K.\ contributed equally to this work.}
\author{Lucas R. Hofer}
\author{Gavin Lamb}
\author{Anna L. Marchant}
\thanks{Present address: STFC Rutherford Appleton Laboratory, Didcot, OX11 0QX, United Kingdom}
\author{Robert P. Smith}
\email{robert.smith@physics.ox.ac.uk}

\affiliation{Clarendon Laboratory, University of Oxford, Parks Road, Oxford, OX1 3PU, United Kingdom}

\newdate{date}{12}{09}{2023}
\date{\displaydate{date}}

\begin{abstract}
	Ultracold gases of highly magnetic lanthanide atoms have enabled the realisation of dipolar quantum droplets and supersolids. However, future studies could be limited by the achievable atom numbers and hindered by high three-body loss rates. Here we study density-dependent atom loss in an ultracold gas of \ce{^166Er} for magnetic fields below \SI{4}{\gauss}, identifying six previously unreported, strongly temperature-dependent features. We find that their positions and widths show a linear temperature dependence up to at least \SI{15}{\micro\kelvin}. In addition, we observe a weak, polarisation-dependent shift of the loss features with the intensity of the light used to optically trap the atoms. This detailed knowledge of the loss landscape allows us to optimise the production of dipolar BECs with more than \num{2e5} atoms and points towards optimal strategies for the study of large-atom-number dipolar gases in the droplet and supersolid regimes.
\end{abstract}

\maketitle

\section{Introduction}
Precise knowledge and control of the nature and strength of interparticle interactions have been a key factor in the success of using degenerate ultracold-atom samples for studying many-body quantum phenomena. The application of a magnetic field close to a Feshbach resonance is a highly versatile and convenient tool for tuning the sign and strength of $s$-wave contact interactions that typically dominate in ultracold gases~\cite{Chin2010}. However, approaching a Feshbach resonance also leads to the enhancement of (detrimental) three-body processes, which result in atom loss and heating~\cite{Weber2003}. Knowing the location of Feshbach resonances and quantifying the associated loss features is thus essential for designing and optimising ultracold-atom experiments.

The realisation of ultracold samples of highly magnetic erbium~\cite{Aikawa2012} and dysprosium atoms~\cite{Lu2011}, which interact via both long-range, anisotropic dipole--dipole interactions and tuneable contact interactions, has led to the discovery of dipolar quantum droplets~\cite{Kadau2016,Ferrier-Barbut2016,Chomaz2016} and a supersolid phase~\cite{Bottcher2019,Tanzi2019,Chomaz2019}, which simultaneously exhibits a global phase order and a spontaneous spatial density modulation. While these first experiments were carried out in cigar-shaped traps leading to (relatively simple) one-dimensional~(1D) spatial ordering, more recently droplet arrays and supersolids with two-dimensional~(2D) ordering have also been observed~\cite{Norcia2021,Bland2022}. Theoretical works predict a plethora of novel patterns in 2D systems, including so-called honeycomb, labyrinthine and pumpkin phases~\cite{Baillie2018,Zhang2021,Hertkorn2021,Hertkorn2021a,Poli2021}. However, reaching these exotic states requires degenerate samples with higher atom numbers than those used in these experiments so far (\num{1.4e5}~\cite{Chomaz2019}).

The maximal achievable atom number in an experiment is often restricted by three-body loss processes, which limit the efficiency of evaporative cooling close to degeneracy and can greatly reduce the gas lifetime at (or while approaching) the desired $s$-wave scattering length. Moreover, in order to map out the parameter space of exotic dipolar phases, one needs to tune the relative strength of the contact and dipole--dipole interactions by controlling the strength of the magnetic field. The precise knowledge of the loss landscape as a function of the field strength is therefore paramount. Here we carefully characterise three-body loss in \ce{^166Er} for magnetic fields below \SI{4}{\gauss}, revealing the presence of six previously unreported resonant loss features which display a strong temperature dependence. In light of this, we describe our optimised procedure for the production of \ce{^166Er} Bose--Einstein condensates~(BECs), containing more than \num{2e5} atoms.

\section{Three-body loss Measurements}
\label{sec:3body}

In alkali atoms the (number) density of Feshbach resonances is typically between \SI{0.01}{\per\gauss} and \SI{0.1}{\per\gauss}~\cite{Chin2010}. However, in magnetic lanthanides, including erbium and dysprosium, the anisotropy of the van der Waals and the dipole--dipole interaction potentials leads to coupling between many scattering channels and consequently to an abundance of Feshbach resonances~\cite{Frisch2014,Maier2015, Baumann2014, Khlebnikov2019}, some of which show a strong temperature dependence~\cite{Beaufils2009, Maier2015, Khlebnikov2021}. Here we focus on \ce{^166Er} for magnetic fields below \SI{4}{\gauss}, where Feshbach resonances and associated loss features have been reported at \SI{0.02(5)}{\gauss}, \SI{3.04(5)}{\gauss} and \SI{4.028}{\gauss}~\cite{Patscheider2022}.

\begin{figure*}
	\centering
	\phantomsubfloat{\label{fig:Nvst}}
	\phantomsubfloat{\label{fig:Tvst}}
	\phantomsubfloat{\label{fig:L3vsB}}
	\setlength{\tabcolsep}{0pt}
	\begin{tabular}[c]{p{0.25\textwidth} p{0.75\textwidth}}
		\centering
		\parbox[c]{0.25\textwidth}{\raggedleft\includegraphics[width=0.25\textwidth]{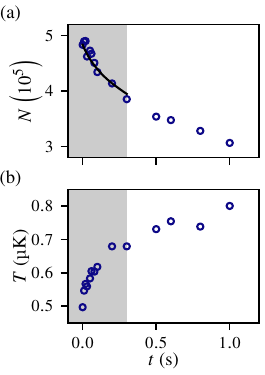}}&
		\centering
		\parbox[c]{0.75\textwidth}{\raggedright\includegraphics[width=0.75\textwidth]{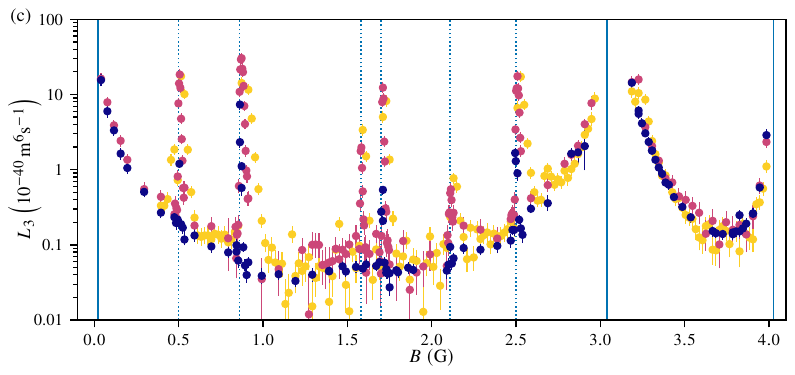}}\\
	\end{tabular}
	\caption{Three-body loss in \ce{^166Er}. \,(a, b)~Time evolution of the atom number~$N$ and temperature~$T$, respectively, of a thermal cloud of \ce{^166Er} atoms at $B=\SI{2.7}{\gauss}$ with an initial temperature of $T_i=\SI{0.5}{\micro\kelvin}$. The shading denotes the region in which $T$ is within \SI{40}{\percent} of $T_i$; in this region we fit $N(t)$ based on \cref{eq:3Bloss_th} to determine $L_3$ (solid line, see text). \,(c)~Extracted three-body loss coefficients~($L_3$) for the different initial temperatures $T_i=\{0.5, 1.5, 4\}\,\unit{\micro\kelvin}$ (blue, magenta and yellow points, respectively). In addition to the three previously identified Feshbach resonances (solid vertical lines), we identify six new loss features (dotted vertical lines). These new features show a noticeable temperature dependence; as $T_i$ increases, the peaks shift to higher $B$ and their widths also increase.}
	\label{fig:L3}
\end{figure*}

\begin{figure*}
	\phantomsubfloat{\label{fig:temp_dep_resonance}}
	\phantomsubfloat{\label{fig:temp_dep_width}}
	\phantomsubfloat{\label{fig:temp_dep_B0}}
	\phantomsubfloat{\label{fig:light_shift_parallel}}
	\phantomsubfloat{\label{fig:light_shift_perpendicular}}
	\centering
	\includegraphics{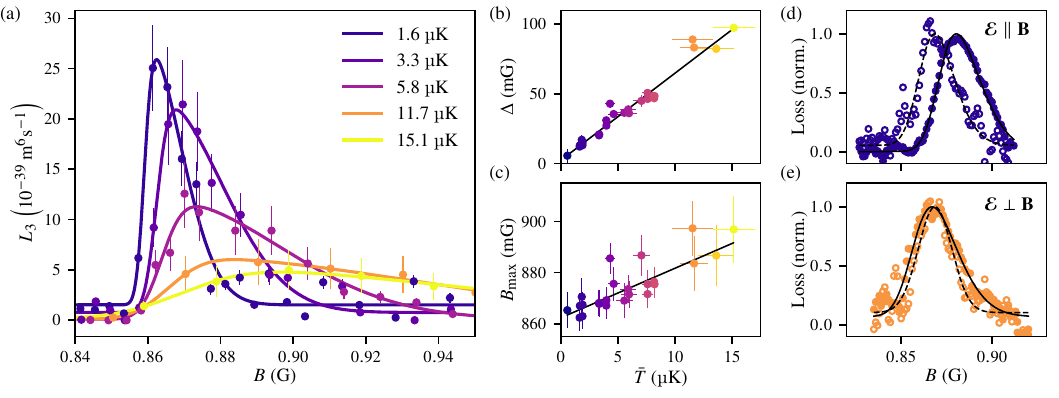}
	\caption{Temperature dependence and light-shift of the loss feature at $\approx \SI{0.86}{\gauss}$. \,(a)~$L_3$ as a function of~$B$ for several decay series with different $\bar{T}$. Markers are experimental data points (with $L_3$ extracted in the same way as in \cref{fig:L3}) and lines are skewed Gaussian fits [see \cref{eq:skewed_gaussian}]. \,(b)~Peak width~$\Delta$ as a function of~$\bar{T}$. The line is a linear fit to the data and has a slope of \SI{6.2(3)}{\milli\gauss\per\micro\kelvin} and an intercept which is consistent with zero. \,(c)~Peak position as a function of $\bar{T}$. The linear fit (line) has a slope of \SI{2.0(3)}{\milli\gauss\per\micro\kelvin} and an intercept of \SI{861(1)}{\milli\gauss}. \,(d,~e)~Light-shift of the loss resonance for linearly polarised light with the polarisation vector $\bm{\mathcal{E}}$ parallel and perpendicular to the dipole orientation, respectively. For both, we plot the normalised loss (see text) for clouds that are prepared at the same temperature~(\SI{2}{\micro\kelvin}) in \SI{1030}{\nano\meter} single-beam ODTs with a factor of four difference in laser intensity (filled points show the higher intensity); for $\bm{\mathcal{E}} \parallel \mathbf{B}$ there is a noticeable light-shift in the resonance position, while the magnitude and width of the feature remains unchanged . Note that the data in (b) and~(c) were corrected for this light-shift effect.}
    \label{fig:resonance}
\end{figure*}

For our measurements we prepare an ultracold, spin-polarised sample of \ce{^166Er} in an (approximately harmonic) optical dipole trap~(ODT) formed from \SI{1030}{\nano\metre} laser light. The experimental sequence is described in \cref{sec:bec}, and the trap and gas parameters for all our loss measurements are given in the Supplemental Material~\footnote{See Supplemental Material for ODT powers, polarisations and associated trap frequencies, the average temperatures and initial atom numbers for all the data series used in \cref{fig:L3,fig:resonance}}; here we only note that the final stage of cooling is achieved by evaporation in the ODT, with the temperature of the atom cloud controlled by the ODT depth. To produce clouds at different temperatures, we interrupt the normal evaporation sequence at different times and ramp up the depth of the ODT over \SI{100}{\milli\second} to prevent any further evaporative cooling (and associated atom loss) during our measurements. We initiate the loss measurements by quenching the magnetic field~$B$~\footnote{The magnetic field is calibrated to an accuracy of \SI{1}{\milli\gauss} using radio frequency (RF) spectroscopy within the ground state Zeeman manifold. However, the quoted magnetic fields are also affected by an error of $\pm\SI{5}{\milli\gauss}$ between different datasets, due to long-term drifts between RF measurements.} to the desired value in~$<\SI{10}{\milli\second}$. To avoid ramping through wide resonances, for measurements above~\SI{3}{\gauss} we evaporatively cool at~\SI{3.8}{\gauss}, whereas for measurements below \SI{3}{\gauss} we cool at~\SI{1.4}{\gauss}. We use absorption imaging after a time-of-flight to measure the atom number~$N$ and temperature~$T$ as a function of the time~$t$ the atoms are held in the trap (at a given~$B$). Examples of these $N(t)$ and $T(t)$ curves are shown in \cref{fig:Nvst,fig:Tvst}; here the initial temperature $T_i=\SI{0.5}{\micro\kelvin}$ and $B=\SI{2.7}{\gauss}$.

Let us first consider atom loss. As the atoms are prepared in the lowest Zeeman state at temperatures much lower than the sub-level splitting ($\approx\SI{78}{\micro\kelvin\per\gauss}$), two-body (spin relaxation) loss processes are energetically suppressed. The evolution of the atom number density in thermal samples can therefore be described by a combination of one- and three-body loss terms~\cite{Weber2003},
\begin{equation}
	\label{eq:3Bloss}
	\dot{n}(\vb{r}) = -\frac{n(\vb{r})}{\tau_1} - L_3 n^3(\vb{r}) \,,
\end{equation}
where $\tau_1$ is the one-body lifetime (set by e.g.\ collisions with background gas atoms in an imperfect vacuum), $L_3$ is the three-body loss coefficient and $n(\vb{r})$ is the atom number density. For a thermal cloud (well above the BEC transition temperature), the atomic density distribution in a harmonic trap is Gaussian and \cref{eq:3Bloss} can be written as~\cite{Weber2003}
\begin{equation}
	\label{eq:3Bloss_th}
		\frac{\dot{N}}{N} = -\frac{1}{\tau_1} - L_3 \left( \frac{m \bar{\omega}^2}{2 \sqrt{3} \pi k_B T} \right)^3 N^2 ,
\end{equation}
where $m$ is the atomic mass, $\bar{\omega}$ is the geometric mean of the trapping frequencies and $k_B$ is the Boltzmann constant. The trapping frequencies were measured separately by exciting the cloud centre-of-mass oscillations in the three perpendicular directions and $\tau_1$ was independently determined to be $\tau_1 = \SI{33(1)}{\second}$ from measurements of low-density clouds for which three-body loss is negligible.

To determine $L_3(B)$ from our $N(t)$ measurements, we fit the numerical solution of \cref{eq:3Bloss_th} to our data [see solid line in \cref{fig:Nvst}] using the corresponding measured $T(t)$ as an input. We only fit our data within the time interval over which the temperature stays within \SI{40}{\percent} of its initial value [gray shaded region in \cref{fig:Nvst,fig:Tvst}] to limit any systematic errors arising from either (i)~evaporative atom loss due to the finite trap depth or (ii)~the fact that $L_3$ may depend on $T$~\footnote{The \SI{40}{\percent} cutoff is chosen as a tradeoff between minimising systematic errors (with a lower cutoff) and random errors (by choosing a higher cutoff to include more data).}.

Regarding the heating of the atom cloud [\cref{fig:Tvst}], this can be understood to be due to two main processes. First, the loss rate is higher in the central (higher density) part of the trap, preferentially removing atoms with energy lower than the average energy in the cloud, leading to `anti-evaporation'~\cite{Weber2003}. Second, the products of the three-body collision can have significant kinetic energy (acquired due to the released binding energy when two atoms form a molecule), which may be partially deposited in the cloud via secondary collisions. 

\Cref{fig:L3vsB} shows the measured three-body coefficient as a function of the magnetic field for initial temperatures of \SI{0.5}{\micro\kelvin}, \SI{1.5}{\micro\kelvin} and \SI{4}{\micro\kelvin}. In addition to the Feshbach resonances already reported [solid vertical lines in \cref{fig:L3vsB}], we observed six additional loss features (dotted vertical lines). These loss features both broaden and shift to higher $B$ with increasing temperature. We note that at temperatures below $\SI{1}{\micro\kelvin}$, where most previous measurements were performed~\cite{Patscheider2022,Frisch2014}, these features become very narrow and can easily be missed.

To explore the temperature dependence further, we measured $L_3$ as a function of $B$ around the newly discovered resonance at $\approx \SI{0.86}{\gauss}$ for several additional $T_i$ values [see \cref{fig:temp_dep_resonance,fig:temp_dep_width,fig:temp_dep_B0}]. Given the asymmetric shape of the loss features, for each $T_i$ data series, $L_3(B)$ was fitted with a heuristic skewed Gaussian curve of the form
\begin{equation}
	\label{eq:skewed_gaussian}
	L_3(B) = A e^{-\frac{(B-B_c)^2}{2\sigma^2}} \left(1 + \erf \mleft( \frac{\alpha(B - B_c)}{\sqrt{2} \sigma}\mright)\right) + C \,,
\end{equation}
where $B_c$, $\sigma$, $\alpha$, $A$ and~$C$ are fitting parameters.

\Cref{fig:temp_dep_width,fig:temp_dep_B0} show, respectively, the peak width~$\Delta$ (taken as twice the variance of the skewed Gaussian) and $B_\text{max}$ [the location of the maximum of $L_3(B)$] as a function of the average temperature~$\bar{T}$ of the decay series~\footnote{Note that the average temperature of a decay series $\bar{T}$ is up to \SI{20}{\percent} higher than the $T_i$ for the same series due to the heating associated with the three-body loss.}. We observe that $\Delta$ grows linearly with temperature and so we parameterise the width of the resonance via a linear function, $\Delta=\Delta_0+(\tdv{\Delta}{T}) \,\bar{T}$, fitted to the data [solid line, \cref{fig:temp_dep_width}]. Note that all our extracted $\Delta_0$ values are consistent with zero within our $\pm \SI{3}{\milli\gauss}$ error bounds. Similarly, $B_\text{max}$ also grows (approximately) linearly with temperature and so we fit the data using $B_\text{max} = B_0+(\tdv{B_\text{max}}{T}) \,\bar{T}$. The parameters of both these fits are tabulated in \cref{tab:Fesh} for all the newly detected loss features. We also note that for the $\SI{0.86}{\gauss}$ feature the maximum $L_3$ decreases with increasing temperature within our measured range [see \cref{fig:temp_dep_resonance}], however, for other peaks this trend is inconclusive.

\begin{table}[b]
	\centering
	\ra{1.3}
	\begin{tabular}{c c c c c}
		\toprule
		$B_0$ & $ \dv{B_\text{max}}{T}$ & $\dv{\Delta}{T}$ & $\mleft(\dv{B_0}{I}\mright)_{\bm{\mathcal{E}} \parallel \vb{B}}$ & $\mleft(\dv{B_0}{I}\mright)_{\bm{\mathcal{E}} \perp \vb{B}}$ \\
		(\unit{\milli\gauss}) & (\unit{\milli\gauss\per\micro\kelvin}) & (\unit{\milli\gauss\per\micro\kelvin}) & (\unit{\gauss\micro\metre\squared\per\watt}) & (\unit{\gauss\micro\metre\squared\per\watt})\\
		\midrule
		498(2) & 3.0(4) & 8.6(2.2) & 1.5(3) & -0.5(2) \\
		862(2) & 1.9(3) & 6.2(3) & 1.7(5) & -0.2(2) \\
		1571(4) & 4.3(7) & 4.6(1.5) & 0.7(2) & 0.1(3) \\
		1705(3) & 2.7(7) & 5.3(1.0) & 1.2(6) & -0.1(3) \\
		2102(3) & 5.6(7) & 7.8(8) & 2.6(9) & 0.1(9)\\
		2497(7) & 3.8(1.6) & 6.1(5) & 1.9(3) & 0.0(4)\\
		\bottomrule
	\end{tabular}
	\caption{Newly detected loss features and their properties. The position of the resonance~$B_0$ for $T \to 0$ (and $I \to 0$) and the rate of change of the peaks' position~($\tdv{B_\text{max}}{T}$) and width~($\tdv{\Delta}{T}$) with temperature, obtained via a linear fit to the finite-temperature datasets. The shift in the resonance position with the intensity of linearly polarised \SI{1030}{\nano\meter} laser light~($\tdv{B_0}{I}$) is also tabulated, for dipoles aligned along ($\bm{\mathcal{E}} \parallel \mathbf{B}$) and perpendicular to ($\bm{\mathcal{E}} \perp \mathbf{B}$) the light polarisation.}
	\label{tab:Fesh}
\end{table}

The magnetic field dependence of the loss properties is due to the differential Zeeman shift arising from the difference in magnetic moments ($\delta \mu$) between the different scattering channels. However, it is also possible for light fields to exert similar differential shifts~\cite{Bauer2009,Fu2013,Cetina2015,Clark2015} due to the difference in polarisabilities ($\delta \alpha$) between scattering channels, which can, in some cases, also have vectorial and tensorial parts~\cite{Becher2018, Chalopin2018}. To check if the optical field from our ODT causes such an effect, we measured the loss features for thermal clouds at the same temperature~(\SI{2}{\micro\kelvin}) but for traps with two different light intensities (powers) and for the polarisation of the ODT light~$\bm{\mathcal{E}}$ either parallel ($\bm{\mathcal{E}} \parallel \vb{B}$) or perpendicular ($\bm{\mathcal{E}} \perp \vb{B}$) to the external magnetic field (and hence the spin-polarisation of the atoms)~\footnote{Note that for linearly polarised light, there is no vector component of the polarisability~\cite{Becher2018} and so the angle between $\mathbf{B}$ and the direction of propagation of the light does not matter.}. Data for the $\SI{0.86}{\gauss}$ resonance is shown in \cref{fig:light_shift_parallel,fig:light_shift_perpendicular}; here, to identify the peak position we simply performed a two-point loss measurement~\footnote{In our two-point measurements the loss is $N(0)-N(t_\text{hold})$, where $t_\text{hold}$ is a fixed hold time, the peak loss is then normalised to 1.}. For $\bm{\mathcal{E}} \parallel \vb{B}$ we observe a significant (positive) shift of the loss feature with light intensity, whereas for $\bm{\mathcal{E}} \perp \vb{B}$ the effect is much less noticeable and (if anything) has the opposite sign. Assuming that the resonance position shifts linearly with light intensity, one can extract a constant of proportionality between the light intensity~$I$ and the resonance peak shift which gives $\tdv{B_0}{I}$ for both orientations. These are tabulated in \cref{tab:Fesh} for all the newly detected loss features.

We note that, for all the loss features, the fact that the difference in the slopes $\tdv{B_0}{I}$ between the two orthogonal light polarisations is of similar magnitude to either of the individual slopes suggests that the tensorial part of $\delta\alpha$ is of a similar magnitude to its scalar part. More quantitatively, using $\delta \alpha = 2 \varepsilon_0 c \delta \mu\, \tdv{B_0}{I}$, we find $\delta \alpha \sim \SI{1}{\atomicunit}$ using $\delta\mu \sim \mu_B$, which is the same order of magnitude as the tensorial part of the ground state polarisability at the wavelength of our ODT~\cite{Becher2018}.

We now compare our findings to the predictions of a `resonant trimer' model previously proposed in the context of temperature-dependent loss features in the lanthanides~\cite{Maier2015}. In this model the loss features are caused by resonances with `closed channel' three-atom (trimer) bound states. We note that as the resonance is with a trimer (rather than a two-atom bound state as in more conventional Feshbach resonances), one would not expect these resonances to affect the (two-body) $s$-wave scattering length~\footnote{Consistent with this, we note that we found no significant change of the cross-thermalisation rate across the \SI{0.86}{\gauss} resonance}.

Within this model, some simple scalings emerge for $k_B T \gg \Gamma_\text{br} \gg \Gamma(E)$, where $\Gamma_\text{br}$ is the trimer decay rate (into an atom and dimer pair) and $\Gamma(E)$ is the collision energy dependent width of the trimer resonance. In this regime, one finds that $B_\text{max} - B_0 =(\lambda+2) k_B T/\delta\mu$, $\Delta =2\sqrt{3+\lambda} k_B T/ \delta\mu$ and $L_3(B_\text{max}) \propto T^{\lambda-1}$, where $\lambda$ is related to the orbital angular momentum of the entrance channel ($\lambda=0, 2$ for $s$-wave and $d$-wave respectively).

The first thing to note is the qualitative agreement between these predictions and our observed temperature dependencies of $B_\text{max}$ and $\Delta$. More quantitatively, the ratio of $\tdv{\Delta}{T}$ and $\tdv{B_\text{max}}{T}$ is predicted to only depend on $\lambda$ and be equal to 1.73 and 1.12 for $\lambda=0$ and 2, respectively. For the \SI{0.86}{\gauss} resonance this ratio is 3.2(5), which is closer to the $s$-wave prediction~\footnote{Note that as the assumption $\Gamma_\text{br} \gg \Gamma(E)$ weakens, we numerically found that while the linear $T$-dependence of $B_\text{max}$ and $\Delta$ still approximately hold, the ratio of the gradients becomes larger.}. An $s$-wave assignment would also be consistent with the fact that the maximum $L_3$ drops with temperature. For the other resonances, the larger error bars and the inconclusive trends of $L_3(B_\text{max})$ against $T$ make any $\lambda$ assignment difficult.

Finally, we note that, despite the consistency of the \SI{0.86}{\gauss} feature with the resonant trimer model, we cannot rule out alternative models~\cite{Beaufils2009} which also predict regimes with linear $T$-dependence of both $B_\text{max}$ and $\Delta$, especially for the other less well-mapped loss features.

\section{Optimised BEC production}
\label{sec:bec}

To produce erbium BECs, we employ standard laser cooling and trapping techniques and then use our knowledge of $L_3(B)$ to inform and optimise the evaporative cooling sequence.

In the initial steps, similarly to Ref.~\cite{Aikawa2012}, an atomic beam emerging from a high-temperature effusion cell oven is slowed down using a Zeeman slower operating on the broad transition at \SI{401}{\nano\metre}. The slow atoms are then loaded into a narrow-line magneto--optical trap~(MOT) operating on the atomic transition at \SI{583}{\nano\metre}. We typically capture \num{e8} atoms after loading the MOT for \SI{12}{\second}. Afterwards, we ramp to a compressed MOT~(cMOT) configuration in \SI{600}{\milli\second}, where reducing the light detuning and intensity causes simultaneous compression and cooling, resulting in a spin-polarised atomic sample at a temperature of \SI{10}{\micro\kelvin}.

\begin{figure}
    \phantomsubfloat{\label{fig:ODTs}}
    \phantomsubfloat{\label{fig:evap_sequence}}
    \phantomsubfloat{\label{fig:evap_efficiency}}
    \centering
    \setlength{\tabcolsep}{4pt}
    \begin{tabular}[c]{p{0.46\textwidth} p{0.46\textwidth}}
	(a) & (b)\\
	\centering
	\parbox[c]{0.46\textwidth}     
        {\raggedleft\includegraphics[width=0.46\textwidth]{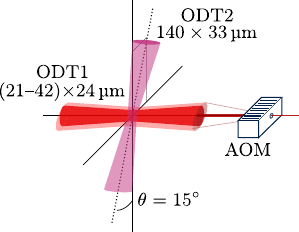}} & \multirow{3}{0.46\textwidth}[52.5pt]{	
            \centering
		\parbox[c]{0.46\textwidth}            
            {\raggedright\includegraphics[width=0.46\textwidth]{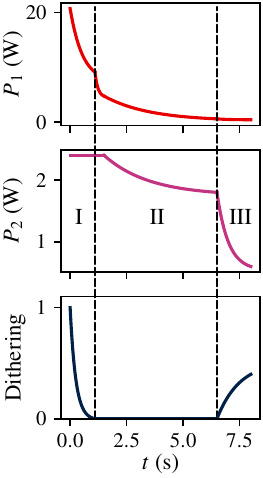}}
        }\\
        (c) & \\
        \centering
	\parbox[c]{0.46\textwidth}         
        {\raggedright\includegraphics[width=0.46\textwidth]{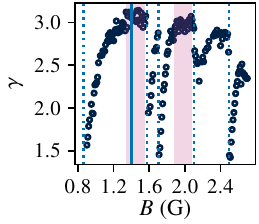}} & \\
    \end{tabular} 
    \caption{Optimising evaporation. \,(a)~Schematics of the crossed-beam optical dipole trap. The horizontal~(ODT1) beam can be enlarged by dithering it with an AOM. The cross~(ODT2) beam propagates at \SI{15}{\degree} to the vertical. \,(b)~After loading the ODT from the cMOT, the evaporation sequence consists of three parts. During phase~I, the power in ODT1 is lowered in parallel with ramping down the dithering. In phase~II, the atoms are drawn into the crossing with the second ODT beam as the cooling continues. Finally, in phase~III, with atoms residing entirely in the crossed region, we decompress the trap by ramping down ODT2 and dithering the ODT1 beam. $P_1$ and $P_2$ correspond to the power in ODT1 and ODT2, respectively, and the dithering amplitude is shown in arbitrary units. \,(c)~Evaporation efficiency~$\gamma$ (see text) as a function of magnetic field. The dashed vertical lines show the $T \to 0$ positions of the loss features. The shaded regions denote optimal regions for evaporation and $B=\SI{1.4}{\gauss}$, which we use during evaporation, is indicated by a solid vertical line.}
\end{figure}

To cool the sample further, we transfer the atoms into an ODT, broadly following previous protocols, in which we perform evaporative cooling. As shown in \cref{fig:ODTs}, the ODT is implemented using two crossed, far-detuned beams at \SI{1030}{\nano\metre}, which we call ODT1 and ODT2. Initially, the $\qtyproduct{21 x 24}{\micro\metre}$ waist ODT1 beam is superimposed onto the cMOT, with a total power of \SI{21}{\watt} and with a \SI{50}{\kilo\hertz} spatial dithering applied using an acousto-optic modulator~(AOM)~\cite{Kohstall2007}, which broadens the horizontal~(\SI{21}{\micro\metre}) waist by a factor of two. A \SI{40}{\milli\second} overlap of the dithered ODT1 beam with the cMOT results in \num{1.8e7} atoms being trapped in ODT1 at a temperature of $\approx \SI{40}{\micro\kelvin}$.

The next step is to evaporatively cool the atoms; it is here where a consideration of three-body losses becomes important. Maximising the evaporation efficiency requires minimising losses while still maintaining a sufficient rate of elastic two-body collisions which facilitate evaporation. To minimise three-body losses in the range of temperatures encountered during evaporation, one wants to be roughly in the middle of the range between the two (relatively broad) Feshbach resonances at \SI{0}{\gauss} and \SI{3}{\gauss}. Additionally, one wants to be as far as possible to the high-field side of any of the loss features, as they both move and broaden towards higher $B$ as $T$ increases. This points towards choosing $B\approx \SI{1.5}{\gauss}$, on the right of the largest gap between loss features [see \cref{fig:L3vsB}]. On the other hand, the elastic collision rate, set by the $s$-wave scattering length $a_s$, increases as one approaches the \SI{3}{\gauss} resonance from below~\cite{Patscheider2022}. This favours higher $B$ and points towards the regions around \SI{2}{\gauss} and \SI{2.4}{\gauss}.

To discern the optimal $B$ for evaporation, in \cref{fig:evap_efficiency} we plot the efficiency ($\gamma$) of the evaporation ramp down to the point just above condensation as a function of $B$. Here $\gamma=-\mathrm{d}(\ln{\rho_0})/\mathrm{d}(\ln{N})$, where $\rho_0=n_0\lambda_T^3$ is the peak phase-space density with $\lambda_T=\sqrt{{2 \pi \hbar^2}/{m k_B T}}$ the thermal de Broglie wavelength. We see that there is indeed an optimal region around $B=\SI{1.4}{\gauss}$ and so we perform our evaporation there, at which point $a_s=73 a_0$~\cite{Patscheider2022}. We note that the region around \SI{2}{\gauss} is also suitable for evaporation, as although it suffers from greater three-body loss, it has a larger $a_s \approx 80 a_0$. This observation is consistent with the \SI{1.9}{\gauss} field used previously for evaporation (see the supplemental material of Ref.~\cite{Chomaz2019}).

\begin{figure}[t]
	\centering
	\phantomsubfloat{\label{fig:psd_evol}}
	\phantomsubfloat{\label{fig:530img}}
	\phantomsubfloat{\label{fig:530fit}}
	\phantomsubfloat{\label{fig:360img}}
	\phantomsubfloat{\label{fig:360fit}}
	\phantomsubfloat{\label{fig:140img}}
	\phantomsubfloat{\label{fig:140fit}}
	\includegraphics{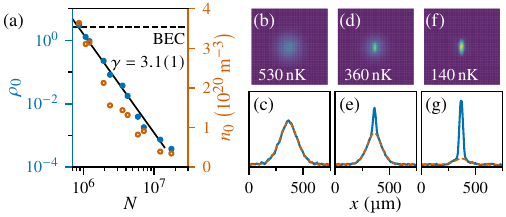}
	\caption{Evaporation to BEC. \,(a)~Evolution of the peak phase-space density~($\rho_0$, blue circles) and number density~($n_0$, orange circles) with the total atom number~($N$) during evaporation. A linear fit on the logarithmic plot shows that $\gamma = -\mathrm{d}(\ln{\rho_0})/\mathrm{d}(\ln{N})=3.1(1)$. \,(b--f)~Time-of-flight absorption images and summed density plots of atomic clouds: (b,\:c)~just prior to condensation, corresponding to the highest $\rho_0$ point in (a), \,(d,\:e)~a partially condensed cloud, \,(f,\:g)~a nearly pure BEC with \num{2.2e5} atoms in the condensate. The dashed line in the summed density plots represents a fit of the extended Bose distribution to the thermal component of the cloud, from which the temperature can be extracted.}
	% In case someone needs this: t_tof is 18ms for (b) and (d), and 24ms for (f)
\label{fig:psd}
\end{figure}

The evaporation sequence can be split into three stages [see \cref{fig:evap_sequence}]. In stage~I, in which the ODT2 contributes negligibly to the trapping, we simultaneously reduce the ODT1 power and ramp down its dithering. This leads to evaporation and a change in the trap aspect ratio, but avoids too much decompression. At the start of stage~II, the ODT2 beam, with a waist of $\qtyproduct{140 x 33}{\micro\metre}$ and an initial power of \SI{2.4}{\watt}, starts to have a noticeable effect and, as the cooling continues, the remaining atoms converge into the crossing of the ODT beams. In stage~III, we employ the novel approach of broadening the ODT1 beam again by ramping up the dithering amplitude alongside significantly decreasing the power of ODT2. This lowers the trap depth and all trapping frequencies, and thereby reduces the atomic density and hence the rate of inelastic three-body collisions relative to the elastic two-body ones.

In \cref{fig:psd_evol} we show how the peak density~$n_0$ and the peak phase-space density~$\rho_0$ evolve with the falling $N$ during the evaporation sequence. This highlights the growing density and justifies the need for our stage~III decompression: at the end of stage~II we reach $n_0=\SI{3e20}{\per\metre\cubed}$ which gives a characteristic three-body lifetime at the centre of the cloud of only $1/L_3 n_0^2 \approx \SI{1}{\second}$. We achieve efficient evaporation throughout the three stages, maintaining a steady increase of $\rho_0$ with efficiency $\gamma=3.1(1)$; this results in the onset of condensation being reached with $N=\num{8e5}$ atoms and at a temperature of \SI{500}{\nano\kelvin}. Finally, by evaporating further we achieve a nearly pure condensate with \num{2.2e5} atoms [see \cref{fig:530img,fig:530fit,fig:360img,fig:360fit,fig:140img,fig:140fit}].

\section{Conclusion}
\label{sec:conclusion}
In conclusion, we have identified six new strongly temperature-dependent three-body loss features in \ce{^166Er} below \SI{4}{\gauss}. Both the position and width of these loss features increase linearly with temperature for $0.5<T<\SI{15}{\micro\kelvin}$; this is broadly consistent with a `resonant trimer' model previously put forward to explain some loss features in lanthanide atoms~\cite{Maier2015}.

Using our knowledge of the loss landscape to optimise the evaporation procedure enabled the production of large BECs of \num{2.2e5} atoms, providing a good starting point for the investigation of ultracold dipolar physics. Furthermore, these findings will enable the optimisation of atom numbers in existing and future experiments, and guide the way towards the experimental realisation of more exotic states, including honeycomb, labyrinthine and pumpkin phases. Moreover, precise knowledge of the three-body loss coefficient could enable the measurement of the atom number density, crucial for determining the structure of quantum droplets.

\begin{acknowledgments}
We thank Nathaniel Vilas for contributions to the early stages of the experiment, and Raphael Lopes and Jean Dalibard for useful discussions. This work was supported by the UK EPSRC (grants no.\ EP/P009565/1 and EP/T019913/1). R.~P.~S.\ and P.~J.\ acknowledge support from the Royal Society, P.~J.\ acknowledges support from the Hungarian National Young Talents Scholarship, M.~K.\ from Trinity College, Cambridge, J.~K.\ from the Oxford Physics Endowment for Graduates~(OXPEG) and G.~L.\ from Wolfson College, Oxford.
\end{acknowledgments}

\bibliography{main}

\end{document}